\documentclass[10pt,twocolumn,twoside]{IEEEtran} 
\ifCLASSINFOpdf
\else
\fi

\usepackage{amssymb}

\usepackage{lineno}
\usepackage[linesnumbered,ruled,vlined]{algorithm2e}
\usepackage{amsmath}
\usepackage{multirow}
\usepackage{graphicx} 
\usepackage{caption} 
\usepackage{subcaption} 
\usepackage{placeins}
\usepackage{amsmath}

\usepackage{placeins}
\usepackage{graphicx} 
\usepackage{caption} 
\usepackage{subcaption} 
\usepackage{pgfplots}
\usepgfplotslibrary{dateplot}

\SetKwInput{KwInput}{Input}

\newcounter{groupcount}
\pgfplotsset{
    draw group line/.style n args={6}{
        after end axis/.append code={
            \setcounter{groupcount}{0}
            \pgfplotstableforeachcolumnelement{#1}\of#6\as\cell{%
                \def\temp{#2}
                \ifx\temp\cell
                    \ifnum\thegroupcount=0
                        \stepcounter{groupcount}
                        \pgfplotstablegetelem{\pgfplotstablerow}{X}\of#6
                        \coordinate [yshift=#4] (startgroup) at (axis cs:\pgfplotsretval,0);
                    \else
                        \pgfplotstablegetelem{\pgfplotstablerow}{X}\of#6
                        \coordinate [yshift=#4] (endgroup) at (axis cs:\pgfplotsretval,0);
                    \fi
                \else
                    \ifnum\thegroupcount=1
                        \setcounter{groupcount}{0}
                        \draw [
                            shorten >=-#5,
                            shorten <=-#5
                        ] (startgroup) -- node [anchor=base, yshift=0.5ex] {#3} (endgroup);
                    \fi
                \fi
            }
            \ifnum\thegroupcount=1
                        \setcounter{groupcount}{0}
                        \draw [
                            shorten >=-#5,
                            shorten <=-#5
                        ] (startgroup) -- node [anchor=base, yshift=0.5ex] {#3} (endgroup);
            \fi
        }
    }
}

\begin{document}
\title{Traffic-Aware Service Relocation in Cloud-Oriented Elastic Optical Networks}

\author{R\'o\.za Go\'scie\'n

\thanks{R. Go\'scie\'n is with the Department of Systems and Computer Networks, Faculty of Electronics, Wroclaw University of Science and Technology, Wroclaw, Poland,
e-mail: roza.goscien@pwr.edu.pl.}% <-this % stops a space
}

\maketitle

\begin{abstract}

In this paper, we study problem of efficient service relocation (i.e., changing assigned data center for a~selected client node) in elastic optical networks (\textsc{eon}) in order to increase network performance (measured by the volume of accepted traffic). To this end, we first propose novel traffic model for cloud ready transport networks. The model takes into account four flow types (i.e., city-to-city, city-to-data center, data center-to-data center and data center-to-data center) while the flow characteristics are based on real economical and geographical parameters of the cities related to network nodes. Then, we propose dedicated flow allocation algorithm that can be supported by the service relocation process. We also introduce 21 different relocation policies, which use three types of data for decision making -- network topological characteristics, rejection history and traffic prediction. Eventually, we perform extensive numerical experiments in order to: (\textit{i}) tune proposed optimization approaches and (\textit{ii}) evaluate and compare their efficiency and select the best one. The results of the investigation prove high efficiency of the proposed policies. The propoerly designed relocation policy allowed to allocate up to 3\% more traffic (compared to the allocation without that policy). The results also reveal that the most efficient relocation policy bases its decisions on two types of data simultaneously -- the rejection history and traffic prediction.

\end{abstract}

\begin{IEEEkeywords}

traffic modeling, traffic prediction, service relocation, anycast traffic, data center 

\end{IEEEkeywords}

% \fi
\IEEEpeerreviewmaketitle

\section{Introduction}

\IEEEPARstart{T}{he} telecommunication networks have become an indispensable part of the society every day life providing support for various human activities -- business, education, finances, heath care, entertainment, social life, etc. As a~result, the number of network users and connected devices continuously increases. According to Cisco company \cite{Cisco_air}, there will be 5.3 billion total Internet users (66\% of global population) by 2023, up from 3.9 billion (51\% of global population) in 2018. Alongside we observe huge popularity of applications related to data centers (\textsc{dc}s), which are able to provide remotely network users with access to plethora of various services, platforms or even computing/storage resources. However, the traffic volume generated by these applications is tremendous while its value is expected to further increase. The telecommunication networks have to in turn endlessly develop and evolve in order to meet the users requirements. Therefore, improvements and completely new superior technologies are designed and implemented in network infrastructures. One of these solutions is architecture of elastic optical networks (\textsc{eon}s). Thanks to the operations within flexible frequency grids and support for advanced transmission and modulation techniques~\cite{Goscien_network15}, \textsc{eon}s are able to provide significantly higher performance than their predecessor, i.e., wavelength division multiplexing (\textsc{wdm}) technology, which is still widely deployed. 

Although providing numerous benefits, the adoption of advance transmission techniques and architectures in optical networks has made their design and operation optimization extremely complex, due to the high number of tunable parameters to be  considered \cite{Musumeci_comsurvtut}. For instance, an effective allocation algorithm in \textsc{eon}s may simultaneously takes into account path choice, modulation format selection, channel assignment, spectrum fragmentation, required bit error rate at the receiver node, survivability provisioning, etc \cite{Goscien_network15, Musumeci_comsurvtut, Klinkowski_transnet19}. Therefore, an improvement in the field of the network design and optimization approaches is required. On the one hand it might be realized by revisiting and enhancing existing methods, on the other hand -- some completely new techniques may be utilized. The networks by themselves have become essential sources of data relevant for their efficient optimization. The modern optical networks are equipped with a~large number of monitors, which are able to gather and provide several types of information on the entire network system. The information can relate to the current and historical resource utilization, traffic volume and profile, failure accidents, quality of services, devices status, etc. As a~result, the attention of many researchers and practitioners has been recently attracted by advanced modeling methods and application of various mathematical approaches derived from machine learning (\textsc{ml}) discipline \cite{Musumeci_comsurvtut}. The advanced modeling approaches are applied to model processes essential for network design and operation, such as traffic pattern and services' profiles. Then, various \textsc{ml} algorithms might be used to process the network-related data in order to forecast and estimate network state and performance in upcoming time frames. The examples of their successful applications include high-accuracy traffic forecasts, flow classification, predictions of such parameters as signal to noise ratio, bit error rate, optical power levels, etc. \cite{Musumeci_comsurvtut}. 

The provision of data center-related services is almost always realized by means of anycasting defined as one-to-one of many transmission technique. A~particular network service is available in a~number of data centers. In order to guarantee data integrity and coherency between different \textsc{dc}s, they continuously communicate with each other and exchange synchronization data. When a~client node requests for a~service, one data center is selected to serve that client. The \textsc{dc} selection process might be performed with respect to various criteria -- geographical distance, transmission delay, \textsc{dc}s' utilization, etc. In order to simplify network management, typically the \textsc{dc}s assigned to client nodes do not change frequently and do not change at all during traffic provisioning. But when the network state and load change it might be beneficial to relocate some services, i.e., to change \textsc{dc} node assigned for a~selected client node. The efficiency of service relocation was proved for static design of survivable optical grids, in which working and backup light-paths are assigned to different \textsc{dc}s \cite{Develder_trannet14, Silva_jocn16, Goscien_rndm14}. Since the modern networks provide solutions for effective centralized management (for instance software defined networking (\textsc{sdn}) architecture), the relocation might be successfully applied to improve performance of dynamic network operations. 

In this paper, we address problem of efficient service relocation in cloud ready elastic optical networks in order to increase network performance (measured by the volume of accepted traffic). To this end, we first propose novel traffic model for transport networks. The model takes into account four flow types (i.e., city-to-city, city-to-data center, data center-to-data center and data center-to-data center) while the flow characteristics are based on real economical and geographical parameters of the cities related to network nodes. Second, we propose dedicated traffic allocation algorithm that can be supported by the service relocation process. Third, we propose 21 different relocation policies, which benefit from network topological characteristics and various data analytics tools applied to the data gathered during network operation --  flow rejection history and traffic history (which might be used for traffic forecasting). Eventually, we perform extensive numerical experiments in order to: (\textit{i}) tune proposed optimization approaches and (\textit{ii}) evaluate and compare their efficiency and select the best one.

The rest of paper is organized as follows. Section 2 reviews related works. Section 3 introduced the traffic model. Section~4 defines optimization problem. Sections 5 and 6 propose dedicated flow allocation algorithm and service restoration policies. The results of numerical experiments are reported in Section 7. Eventually, Section 8 concludes the paper. 

\section{Related works}

In order to increase the results valence and make them applicable in real networks, it is crucial to use in numerical experiments models and/or datasets reflecting realistic patterns observed in the network traffic. 

There are publicly available numerous datasets with historical data related to traffic observed at some specific network links/nodes. For instance, \textsc{sndlib} library \cite{orlowski_inoc07} offers traffic matrices defined for static and dynamic scenarios for several real network topologies. The data was provided by network operators and researchers. The main drawback of the library is fact that the majority of the data was gathered before 2014. Then, Seattle Exchange Point (\textsc{six})~\cite{six} shares history of incoming/outgoing bit-rates (within given time window) at routers located in the \textsc{six}. Similarly, several other platforms publish general information regarding observed traffic at some Internet exchange points. For example, Amsterdam Internet Exchange~\cite{aix} shares data from Amsterdam. There is also available information regarding traffic in different locations such as University Campus of \textsc{agh} University of Science and Technology (Krakow, Poland)~\cite{Jurkiewicz_comcom21} or Wroclaw Center for Networking and Supercomputing (Wroclaw, Poland)~\cite{wcss}. The mentioned datasets provide tremendous amount of historical data that can be used for research especially concerning traffic prediction. However, the main drawback of these datasets is fact that, due to the privacy and security reasons, they show only selected characteristics of the traffic. For instance, aggregated bit-rate over time or number of packets. Moreover, they share the incoming/outgoing traffic only for selected exchange point/links. Therefore, they cannot be directly applied to simulate traffic in multi-node networks and do not allow to study realization of various network services and the related traffic patterns. 

To overcome problems and limitations related to historical datasets, the researchers use various traffic generation models in their experiments. One of the oldest and most commonly applied model assumes that traffic demands arrive to the network following a~Poisson  process while their lifetime is generated according to a~negative exponential distribution~\cite{Ba_transnet17, Goscien_rndm19, Walkowiak_jocn18}. But the model is based on the assumptions made for the traditional telephony networks and, therefore, it does not reflect patterns and characteristics observed in the convergent optical networks supporting plethora of diversified services. Therefore, novel models are required. Ref.~\cite{Gaizi_issc16} suggests to model traffic in wavelength division multiplexing (\textsc{wdm}) networks using Pareto process and shows its successful application. The authors of~\cite{Gencata_trannet03, Troia_osn20} elaborate network-dedicated models based on the gathered observations of the traffic within specific time window. Ref.~\cite{Vela_icton16} proposes realistic traffic functions that can be used for the modeling purposes. The proposals include piecewise linear function with mean value following the Gaussian distribution, sine function and the combination of two first options. The paper gives general functions formulas and does not precisely define their parameters. Then, the authors of~\cite{Goscien_network15, Klinkowski_network13} apply the multivariable gravity model where the bit-rate exchanged between a~pair of nodes is determined by real data related to the populations of the regions served by the network nodes, geographical distance between the nodes and economy level expressed by gross domestic product (\textsc{gdp}). It is worth-mentioning that due to the various limitations none of the proposed models was widely accepted and applied in the research society. Hence, a~lot of effort is still required in the field of traffic modeling.

Due to the increasing systems' complexity and data availability, the computational intelligence (and especially machine learning) has been identified as a~promising solution to improve network performance. One of the most popular applications of \textsc{ml} algorithms in the field of networking is the problem of traffic prediction~\cite{Musumeci_comsurvtut}. For that purpose, the most commonly applied methods are autoregressive integrated moving average (\textsc{arima})~\cite{Fernandez_jocn15, Morales_jocn17} and various implementations of neural networks (\textsc{nn})~\cite{Troia_icton18, Balanici_ofc19, Singh_jocn18}. The methods have revealed high prediction accuracy for the traffic forecasting task. The efficient traffic prediction tool might be then applied for direct network optimization. For instance, Ref.~\cite{Troia_icton18} uses forecasted data in order to solve routing, modulation and spectrum assignment (\textsc{rmsa}) problem in \textsc{eon}s. The manuscript~\cite{Singh_jocn18} applies prediction of traffic features to propose efficient resource (re)allocation strategy in optical data center networks. Then, the authors of~\cite{Eramo_icton20} developed a~traffic forecasting algorithm for the resource allocation in network function virtualization (\textsc{nfv}) network architectures in which data centers are interconnected by an \textsc{eon}. Their procedure aims to forecast the traffic so as to minimize the network operation cost. Eventually, Refs.~\cite{Fernandez_jocn15, Morales_jocn17} study problem of efficient adaptation / reconfiguration of virtual network topologies (\textsc{vnt}s), which were introduced to improve resource management and provision by network operators. 

The service relocation process in optical networks was widely explored in the literature in the context of service protection/restoration after a~failure~\cite{Develder_trannet14, Silva_jocn16, Goscien_rndm14}. The authors of~\cite{Develder_trannet14} investigate static optical grid dimensioning problem, where the required protection level is guaranteed by shared backup path protection (\textsc{sbpp}) scheme. Each service is provided here by two light-paths, which can relate to different \textsc{dc} nodes. Ref.~\cite{Goscien_rndm14} addresses similar problem in \textsc{eon}s, wherein the protection is provided by dedicated path protection (\textsc{dpp}). Both studies prove high efficiency of the service relocation process in the static network design problem. Then, paper~\cite{Silva_jocn16} covers flow restoration in optical network. The authors propose the strategy in which the restoration light-paths do not have to use the same \textsc{dc} node as the primary ones. The results of numerical experiments demonstrate that the approach improves the average service availability and restorability performance. In the context of dynamic traffic provisioning, the literature mostly focuses on the efficient \textsc{dc} selection for incoming connections~\cite{Goscien_network15, Klinkowski_network13, Zhang_comlet16}. The decision is based on parameters such as shortest path length (measured in kilometres or number of hops), server load, transmission delay, etc. 

Summarizing, a~lot of effort is still required in the field of efficient traffic modeling in optical networks. Moreover, the literature lacks studies focused on the dynamic service relocation in optical networks especially based on advanced methods such as data analytics. The proposed paper answers the identified important directions of research work and fills the literature gaps. 

\section{Traffic modeling}\label{sec:traffic-model}

In this section, we propose novel traffic model for transport national or international networks. Four traffic types are taken into account -- city-to-city, city-to-\textsc{dc}, \textsc{dc}-to-city and \textsc{dc}-to-\textsc{dc}. City-to-city traffic is observed between each pair of network nodes and represents general traffic between different cities. Then, city-to-\textsc{dc} and \textsc{dc}-to-city are observed for each pair of non-\textsc{dc} node (city) and \textsc{dc} assigned to served that city. City-to-\textsc{dc} reflects service or content requests and transmission control data. \textsc{dc}-to-city traffic concurrently is a~service/content provisioning. The volume and frequency of the transmission from \textsc{dc}-to-city is significantly higher compared to the flow in opposite direction. For a~particular city node, city-to-\textsc{dc} traffic always overtakes \textsc{dc}-to-city traffic in the time domain, since that is a request-response type of a~transmission. Eventually, \textsc{dc}-to-\textsc{dc} traffic occurs between each pair of \textsc{dc} nodes and describes inter-\textsc{dc} synchronisation. The first synchronisation must be performed before city-to-\textsc{dc} / \textsc{dc}-to-city transmission begins. 

For the purpose of modeling, let $i,j, \in V$ indicate network nodes. For each traffic type we use sine (trigonometric) functions to describe data flow between any pair of communicating nodes (i.e., $(i,j): i,j \in V; i \neq j$) at any time point \textit{t}, see eq.~(\ref{eq:general-sinus}). The functions are moved by the amplitude value in order to take only non-negative values. As described in the next subsections, each of the sine functions has its own amplitude, pulsation and initial phase. Additionally, we introduce to the model globally constant value \textit{A}, which represents the maximum signal amplitude. Based on the required network observation time \textit{T} (given in a~number of iterations), \textit{A} value is dynamically assesses to guarantee that average traffic load in each time point is $B_{avg}$ [Tbps]. Please note that \textit{T} and $B_{avg}$  are model's input parameters. 

\begin{equation} \label{eq:general-sinus}
    f(t) = A \cdot a \cdot [\sin{(\omega \cdot t + \phi)}+1]
\end{equation}

Where:
\begin{itemize}
    \item \textit{t} -- a~time stamp.
    \item \textit{f} -- data flow (bit-rate) in Gbps.
    \item \textit{A} -- maximum signal amplitude in Gbps.
    \item \textit{a} -- current signal amplitude.
    \item $\omega$ -- signal pulsation. 
    \item $\phi$ - signal initial phase.
\end{itemize}

The amplitudes, pulsations and initial phases of functions reflecting different traffic types are based on realistic characteristics of cities related to network nodes. In more detail, the model takes into account following characteristics:

\begin{itemize}
    \item \textsc{dist}$(i,j)$ - distance (in km) between nodes $i$ and $j$. 
    \item \textsc{dist}$_{min}$ - minimum distance \textsc{dist}$(i,j)$ between a~pair of nodes $(i,j): i,j \in V; i \neq j$. I.e., \textsc{dist}$_{min} = \min_{i,j \in V, i \neq j} \textsc{dist}(i,j)$.
    \item \textsc{gdp}$(i)$ - gross domestic product (\textsc{gdp}) of the city (region) related to node \textit{i}.
    \item \textsc{pop}$(i)$ - population (measured in millions of citizens) of the city (region) related to node \textit{i}. 
    \item \textsc{gdp}$\_$\textsc{pop}$(i)$ - product of the \textsc{gdp} and population related to node \textit{i}. I.e., \textsc{gdp}$\_$\textsc{pop}$(i) = \textsc{gdp}(i) \cdot \textsc{pop}(i)$.
    \item \textsc{gdp}$\_$\textsc{pop}$_{max}$ - maximum value of \textsc{gdp}$\_$\textsc{pop}$(i)$ among all nodes $i \in V$. I.e., $\textsc{gdp}\_\textsc{pop}_{max} = \max_{i \in V} $
     \textsc{gdp}$\_$\textsc{pop}$(i)$.
     \item $\textsc{gdp}\_\textsc{pop}_{sum}$ - sum of $\textsc{gdp}\_\textsc{pop}(i)$ related to all nodes $i \in V$. I.e., $\textsc{gdp}\_\textsc{pop}_{sum} = \sum_{i \in V} \textsc{gdp}\_\textsc{pop}(i)$.
\end{itemize}

\subsection{City-to-city traffic}

The city-to-city traffic between nodes \textit{i} and \textit{j} is described by eq.~(\ref{eq:city-to-city}). The signal amplitude $a_{cc}(i,j)$ is proportional to the sum of $\textsc{gdp}\_\textsc{pop}$ values obtained for two communicating nodes and inversely proportional to the distance between them. A~special factor $\frac{\textsc{dist}_{min}}{\textsc{gdp}\_\textsc{pop}_{max}}$ is used to keep values of $a_{cc}(i,j)$ in the range of $(0;1)$. The signal pulsation $\omega_{cc}(i,j)$ is also inversely proportional to the distance between communicating nodes and is expressed as the ratio of standard sine period equal to $2\pi$. The signal initial phase $\phi_{cc}(i,j)$ is a~random value selected from the range $<0;\omega_{cc}(i,j)>$.

\begin{equation} \label{eq:city-to-city}
    f_{cc}(t,i,j) = A \cdot a_{cc}(i,j) \cdot [\sin{(\omega_{cc}(i,j) \cdot t + \phi_{cc}(i,j))}+1]
\end{equation}

\begin{equation} \label{eq:amplitude_city-to-city}
    a_{cc}(i,j) = \frac{\textsc{dist}_{min}}{\textsc{gdp}\_\textsc{pop}_{max}} \cdot \frac{ \textsc{gdp}\_\textsc{pop}(i) + \textsc{gdp}\_\textsc{pop}(j) }{\textsc{dist}(i,j) }
\end{equation}

\begin{equation} \label{eq:omega_city-to-city}
    \omega_{cc}(i,j) = 2\pi \cdot \frac{1}{\textsc{dist}(i,j)}
\end{equation}

\subsection{City-to-\textsc{dc} traffic}

The traffic from a~city node \textit{i} to a~\textsc{dc} is given by formula~(\ref{eq:city-to-dc}). Its amplitude $a_{cdc}(i)$ and pulsation $\omega_{cdc}(i)$ are mainly determined by the city $\textsc{gdp}\_\textsc{pop}(i)$ divided by the $\textsc{gdp}\_\textsc{pop}_{max}$. The signal initial phase $\phi_{cdc}(i)$ is a~random value selected from the range $<0;\frac{2\pi}{\omega_{cdc(i)}}>$.
Please note that the traffic formula depends only on the parameters related to a~city node. Therefore, the traffic volume remains the same regardless of a~\textsc{dc} node currently serving the city. 

\begin{equation}\label{eq:city-to-dc}
    f_{cdc}(t,i) = A \cdot a_{cdc}(i) \cdot [\sin{(\omega_{cdc}(i) \cdot t + \phi_{cdc}(i))} +1]
\end{equation}

\begin{equation} \label{eq:amplitude_city-to-dc}
    a_{cdc}(i) = \frac{1}{10} \cdot \frac{\textsc{gdp}\_\textsc{pop}(i)}{\textsc{gdp}\_{\textsc{pop}_{max}}} 
\end{equation}

\begin{equation} \label{eq:omega_city-to-dc}
    \omega_{cdc}(i) = \frac{1}{10} \cdot 2\pi \cdot \frac{ \textsc{gdp}\_\textsc{pop}(i)}{\textsc{gdp}\_\textsc{pop}_{sum}} 
\end{equation}

\subsection{\textsc{dc}-to-city traffic}

The traffic from a~\textsc{dc} to a~particular non-\textsc{dc} node \textit{i} is given by eq.~(\ref{eq:dc-to-city}). Its definition is strongly related to the corresponding traffic in opposite direction (i.e., from the city node \textit{i} to a~\textsc{dc}). In turn, the amplitude $a_{dcc}(i)$ and pulsation $\omega_{dcc}(i)$ are 10 times higher than corresponding parameters defined for the traffic in the opposite direction. The signal initial phase is selected randomly from the range $<\phi_{cdc}(i);2\pi \omega_{dcc}(i)>$.

\begin{equation}\label{eq:dc-to-city}
    f_{dcc}(t,i) = A \cdot a_{dcc}(i) \cdot [\sin{(\omega_{dcc}(i) \cdot t + \phi_{dcc}(i))} +1]
\end{equation}

\begin{equation} \label{eq:amplitude_dc-to-city}
    a_{dcc}(i) = \frac{\textsc{gdp}\_\textsc{pop}(i)}{\textsc{gdp}\_{\textsc{pop}_{max}}} = 10 \cdot a_{cdc}(i)
\end{equation}

\begin{equation} \label{eq:omega_dc-to-city}
    \omega_{dcc}(i) = 2 \pi \cdot \frac{ \textsc{gdp}\_\textsc{pop}(i)}{\textsc{gdp}\_\textsc{pop}_{sum}} = 10 \cdot \omega_{cdc}(i)
\end{equation}

\subsection{\textsc{dc} to \textsc{dc} traffic}

The \textsc{dc} to \textsc{dc} traffic between two \textsc{dc} nodes \textit{i} and \textit{j} is given by the formula (\ref{eq:dc-dc}). In order to express necessity of frequent inter-\textsc{dc}s synchronisation and based on our preliminary experiments, we assume $a_{dcdc}(i,j) = 0.5$, $\omega_{dcdc}(i,j) = 2 \cdot \max_{i,j \in V; i \neq j} \omega_{cc}(i,j)$ and $\phi_{dcdc}(i,j) = 0$.

\begin{equation} \label{eq:dc-dc}
    f_{dcdc}(t,i,j) = A \cdot a_{dcdc}(i,j) \cdot [\sin{(\omega_{dcdc}(i,j) \cdot t + \phi_{dcdc}(i,j))}+1]
\end{equation}

\subsection{Total traffic between a~pair of nodes}

In each time stamp $t \in T$, the total traffic volume between a~pair of nodes \textit{i} and \textit{j} is determined by the sine functions representing traffic types observed between these nodes (see eq.~(\ref{eq:total_traffic})). 

\begin{equation}
\begin{split}\label{eq:total_traffic}
    f(t,i,j) = f_{cc}(t,i,j) + \delta(j) \cdot \gamma(i,j) \cdot f_{cdc}(t,i) + \\ \delta(i) \cdot \gamma(j,i) \cdot f_{dcc}(t,j) + \delta(i) \cdot \delta(j) \cdot f_{dcdc}(t,i,j)
    \end{split}
\end{equation}

Where:
\begin{itemize}
    \item $\delta(i)$ = 1, if node \textit{i} hosts a~\textsc{dc}; 0, otherwise
    \item $\gamma(i,j)$ = 1, if city node \textit{i} is served by \textsc{dc} located in node~\textit{j}; 0, otherwise 
\end{itemize}

Please note that the model assumes a~specific traffic grooming. I.e., in each time stamp $t \in T$ it allows to asses the total bit-rate exchanged between a~pair of nodes \textit{i} and \textit{j}. In order to verify if some new demand (bit-rate) has arrived in time stamp \textit{t}, we need to compare current traffic volume $f(t,i,j)$ with the previous observation $f(t-1,i,j)$. 

\section{Optimization problem}

Formally, the problem covered in this paper is dynamic routing and spectrum allocation (\textsc{rsa}) in elastic optical networks. The objective function is defined as a~bandwidth blocking probability (\textsc{bbp}) and should be minimized. 

\textsc{eon} network is modeled as a~directed graph $G=(V,E)$ where \textit{V} is a~set of network nodes and \textit{E} is a~set of directed fiber links. The spectrum resources available on each link are divided into $S$ frequency slices, which can then grouped into channels $c \in C$. Each channel is characterized by a~first slice index and number of included slices. Moreover, there is a~set~\textit{R} of data centers located in the network. 

The network is observed within \textit{T}-iterations time perspective while the traffic is generated according to the model presented in Section~\ref{sec:traffic-model} assuming average network load to be $B_{avg}$. The model takes into account four traffic types (i.e., city-to-city, city-to-\textsc{dc}, \textsc{dc}-to-city and \textsc{dc}-to-\textsc{dc}) and assesses aggregated traffic volume $f(t,i,j)$ in each time stamp $t \in T$ for each pair of communicating nodes $(i,j): i,j \in V; i \neq j$. In order to realize transmission between pairs city-\textsc{dc}, each client node (i.e., city node) has to be assigned to one of $|R|$ available \textsc{dc}s. The initial assignment is made in the beginning of $t=1$ time stamp. However, it might be changed during network operation by the service relocation process.  

The aim of the problem is to serve as much of the offered bit-rate as it possible within given \textit{T} iterations with respect to the limited network resources and basic \textsc{rsa} constraints (i.e., spectrum continuity, spectrum contiguity, spectrum non-overlapping) \cite{Klinkowski_transnet19}. To serve a~bit-rate between a~pair of nodes $(i,j)$, a~light-path has be established. It is a~combination of a~routing path connecting these nodes and a~channel able to accommodate required bit-rate. The traffic model assesses in each time stamp an aggregated traffic volume for each pair of nodes, however, it might be divided and realized by multiple light-paths. 

To calculate required channel width for a~particular bit-rate and routing path, we apply similar assumptions as in \cite{Khodashenas_lighttech16}. In particular, we assume that a~transponder occupies 3 frequency slices (37.5~GHz) and can use one of four modulation formats: \textsc{bpsk}, \textsc{qpsk}, 8-\textsc{qam}, 16-\textsc{qam}. Table~\ref{tab:modulations} presents supported bit-rate and transmission distance for each modulation. When a~path length exceeds transmission distance supported by a~modulation, we use signal regenerators. To select a~modulation format for a~particular bit-rate on a~candidate routing path, the distance-adaptive transmission (\textsc{dat}) rule is used \cite{Walkowiak_springer16}. It applies the most spectrally efficient format which simultaneously minimizes the number of required regenerators. 
\begin{table}[htb]
    \centering
    \caption{Supported bit-rate and transmission distance for a~transponder operating within 37.5~GHz spectrum \cite{Khodashenas_lighttech16}}
    \label{tab:modulations}
    \begin{footnotesize}
\begin{tabular}{c|cccc}
        
        & \textsc{bpsk} & \textsc{qpsk} &\textsc{ 8-qam} &\textsc{ 16-qam} \\ \hline\hline
       supported bit-rate [Gbps] & 50 & 100 & 150 & 200 \\ 
       transmission reach [km] & 6300 & 3500 & 1200 & 600 \\ 
    
    \end{tabular}
    \end{footnotesize}
\end{table} 

The problem objective function is the bandwidth blocking probability (\textsc{bbp}), which is calculated as follows. For a~particular pair of nodes $(i,j)$ and a~time stamp~\textit{t}, the offered traffic $b_{offered}(t,i,j)$ is given by formula (\ref{eq:offered_traffic}) while the rejected bit-rate $b_{rej}(t,i,j)$ is calculated with respect to that value (i.e., how much of that bit-rate remains unallocated). The offered traffic in a~time stamp \textit{t} is then a~sum of offered traffic for each pair of nodes $(i,j)$, i.e., $b_{offered}(t) = \sum_{(i,j): i,j \in V; i \neq j} b_{offered}(t,i,j)$. Similarly, the traffic rejected in a~time stamp \textit{t} is calculated as a~sum of rejected bit-rate over all pairs of nodes, i.e., $b_{rej}(t) = \sum_{(i,j): i,j \in V; i \neq j} b_{rej}(t,i,j)$. On that background, bandwidth blocking probability $\textsc{bbp}(t)$ in a~time stamp \textit{t} is calculated as a~rejected bit-rate $b_{rej}(t)$ divided by the offered bit-rate $b_{offered}(t)$ (i.e., $\textsc{bbp}(t) = \frac{b_{rej}(t)}{b_{offered}(t)}$). Eventually, \textsc{bbp} withing time period \textit{T }is calculated as average blocking over iterations $t \in T$, i.e., $\textsc{bbp} = \frac{\sum_{t \in T} \textsc{bbp}(t)}{T}$.

\begin{equation}\label{eq:offered_traffic}
    b_{offered}(t,i,j) = \begin{cases}
    a = f(t,i,j) - b_{cur}(i,j), \mbox{ if } 0 \leq  a,  \\
    0,  \mbox{ otherwise }
    \end{cases}
\end{equation}

\section{Routing and spectrum allocation algorithm}

Due to the specific definition of the traffic model, we propose a~dedicated routing and spectrum allocation (\textsc{rsa}) algorithm called Traffic-Dedicated Routing and Spectrum Allocation (\textsc{tdrsa}). The method aims at minimizing bandwidth blocking probability (\textsc{bbp}). To this end, it combines various allocation strategies, tries to minimize number of established light-paths and re-allocation actions. Initially \textsc{tdrsa} assigns each client node with the closest \textsc{dc}. However, to further improve its performance it can be combined with a~service relocation policy. 

For the purpose of the algorithm description, let $p \in P(i,j)$ be a~set of candidate shortest routing paths connecting pair of nodes $(i,j)$. In this study, we measure path length in kilometres. Next, let $l \in L(i,j)$ be a~set of light-paths established between a~pair of nodes $(i,j)$. Each light-path $l = (c,p)$ is a~connection of a~frequency channel \textit{c} and routing path \textit{p} (wherein $p \in P(i,j)$). Depending on the channel size, each light-path is characterized by its maximum capacity $b_{max}(l)$ (i.e., maximum supported bit-rate [Gbps]). Due to the rounding process, the channels are often not fully utilized. Therefore, we use $b_{cur}(l)$ to indicate bit-rate, which is currently transmitted using light-path \textit{l}. Having a~number of established light-paths between a~pair of nodes $(i,j)$, we can assess current bit-rate allocated between that pair as $b_{cur}(i,j) = \sum_{l \in L(i,j)} b_{cur}(l)$. Moreover, in each time stamp~\textit{t}, we can check whether currently offered traffic $f(t,i,j)$ is fully realized (then $ f(t,i,j) \leq b_{cur}(i,j)$).

The idea behind \textsc{tdrsa} method is presented in~Algorithm~\ref{alg:rsa}. In the beginning, it calculates a~set of $\lambda$ shortest paths $P(i,j)$ for each pair of communicating nodes $(i,j), i, j \in V, i \neq j$ (Alg.~\ref{alg:rsa}, line~\ref{alg:ksp}). The paths are calculated using well-known Yen's kSP algorithm. It also assigns each network node (i.e., city node) with the closest \textsc{dc} (Alg.~\ref{alg:rsa}, line \ref{alg:dc-assignment}). Next, the \textsc{tdrsa} simulates network operation within a~period of \textit{T} iterations (Alg.~\ref{alg:rsa}, lines \ref{alg:iter_start}--\ref{alg:iter_stop}). For each time stamp $t \in T$, the method first initializes $b_{rej}(t)$, $b_{total}(t)$ and $\textsc{bbp}(t)$ (Alg.~\ref{alg:rsa}, lines \ref{alg:init_start}--\ref{alg:init_stop}). In the next step, it checks if the condition for service relocation is satisfied (Alg.~\ref{alg:rsa}, line \ref{alg:reloc_con}). If it is, the method runs a~special procedure responsible for adaptive service relocation (see Section~\ref{sec:relocation-policies}). The procedure is run every $\alpha$ iterations and cannot start before $t_{start}$ iterations since the beginning of the network operation. Then, the \textsc{tdrsa} goes through all pairs of communicating nodes $(i,j)$ (Alg.~\ref{alg:rsa}, lines \ref{alg:nodes-start}--\ref{alg:nodes-stop}). If the current flow between a~pair is higher than currently allocated flow for that pair (i.e., $b_{cur}(i,j) < f(t,i,j)$), the values of $b_{offered}(t,i,j)$ and $b_{offered}(t)$ are updated (Alg.~\ref{alg:rsa}, lines \ref{alg:offered-traffic-ij}--\ref{alg:offered-traffic}) and the special function $\textsc{tdrsa\_higher\_flow}(...)$ is run in order to serve offered traffic $b_{offered}(t,i,j)$ (Alg.~\ref{alg:rsa}, line \ref{alg:higher-flow}). If the current flow is lower than currently allocated flow (i.e., $b_{cur}(i,j) > f(t,i,j)$), the method $\textsc{tdrsa\_lower\_flow}(...)$ is run (Alg.~\ref{alg:rsa}, line \ref{alg:lower-flow}). When the flows are equal to each other, no (re-)allocation action is required. Next, the algorithm calculates current blocking probability (i.e., \textsc{bbp}(\textit{t})) and moves to the next iteration  (Alg.~\ref{alg:rsa}, line \ref{alg:tmp-bbp}). After process of \textit{T} iterations, the \textsc{tdrsa} calculates final \textsc{bbp} values (as the average over \textsc{bbp}$(t), t \in T$) and terminates  (Alg.~\ref{alg:rsa}, line \ref{alg:final-bbp}). 

Note that $\lambda$ is $\textsc{tdrsa}$'s input parameters while $\alpha$ and $t_{start}$ are parameters control relocation process. 

\begin{algorithm}
\caption{\textsc{tdrsa} algorithm}\label{alg:rsa}

\begin{small}

\DontPrintSemicolon

\KwInput{ \textit{T}, $G = (V,E)$, $B_{avg}$, $t_{start}$, $\alpha$ }

\For{ \textbf{each} $(i,j): i,j \in V; i \neq j$ }
{
    $P(i,j) \gets kSP(i,j,\lambda)$ \label{alg:ksp}
    
    $L(i,j) \gets \emptyset$
}

\For{ \textbf{each} $i \in V$ }
{
    $\gamma(i,j) = 1, j:$ \textsc{dist}$(i,j) = \min_{j \in V: \delta(j)=1}$ \textsc{dist}$(i,j)$ \label{alg:dc-assignment}
}

\For{$t \in T$}
{ \label{alg:iter_start}
    $b_{rej}(t) = 0$ \label{alg:init_start}
    
    $b_{offered}(t) = 0$
    
    $BBP=0$ \label{alg:init_stop}
    
    \If{($t_{start} \leq t ) \textsc{and} (t \% \alpha = 0)$} 
    { \label{alg:reloc_con}
        \textsc{service\_relocation}$(...)$ \label{alg:relocation}
    }
    
    \For{ \textbf{each} $(i,j): i,j \in V; i \neq j$ }
    {\label{alg:nodes-start}
        $f(t,i,j) \gets assess\_cur\_flow(t,i,j)$ \label{alg:get-flow}
        
        $b_{cur}(i,j) = \sum_{l \in L(i,j)} b_{cur}(l)$ \label{alg:get-cur-flow}
        
        \If{$b_{cur}(i,j) \leq f(t,i,j)$}
        {
        $b_{offered}(t,i,j) = f(t,i,j) - b_{cur}(i,j)$ \label{alg:offered-traffic-ij}
        
        $b_{offered}(t) = b_{offered}(t) + b_{offered}(t,i,j)$ \label{alg:offered-traffic}
        
        \textsc{tdrsa\_higher\_flow}$(...)$ \label{alg:higher-flow}
            %$TDRSATD\_higher\_flow(t,i,j)$
        }
        
        \ElseIf{$b_{cur}(i,j) > f(t,i,j)$}
        {
            \textsc{tdrsa\_lower\_flow}$(...)$ \label{alg:lower-flow}
        }
    }\label{alg:nodes-stop}
    
    $BBP(t) = b_{rej}(t) / b_{offered}(t)$ \label{alg:tmp-bbp}
} \label{alg:iter_stop}

$BBP = \frac{\sum_{t \in T} BBP(t)}{T}$ \label{alg:final-bbp}

\textbf{return} \textsc{bbp}

\end{small}

\end{algorithm}

\subsection{$\textsc{tdrsa\_higher\_flow}(...)$ function}

When the entire flow between a~pair of nodes $(i,j)$ is higher that the flow currently allocated for that pair, the allocation task is solved by function $\textsc{tdrsa\_higher\_flow}(...)$ presented in Alg.~\ref{alg:increased_traffic}. In order to decrease \textsc{bbp} and provide fast on-line calculations, the method tries to serve as much bit-rate as it is possible and simultaneously minimize number of established light-paths and re-allocation actions. In order to allocate offered bit-rate $b_{offered}(t,i,j) = f(t,i,j) - b_{cur}(i,j)$, the method combines five strategies which are always applied subsequently: (\textit{i})~addition to an existing light-path, (\textit{ii}) re-allocation of the entire flow, (\textit{iii}) re-allocation of a~part of the flow, (\textit{iv}) addition to existing light-paths, and (\textit{v}) new light-path establishment.  

In the beginning, the method tries to add offered bit-rate to one of the existing light-paths $l \in L(i,j)$ (Alg.~\ref{alg:increased_traffic}, lines~\ref{alg2:phase1_start}-\ref{alg2:phase1_end}). Please note that the key idea of that strategy assumes that the bit-rate is entirely added to only one of the existing light-paths. The light-paths are considered from the oldest to the newest (according to the time stamp of their establishing). If the addition is possible, the flow is fully served and method terminates. Otherwise, it moves to the second strategy (Alg.~\ref{alg:increased_traffic}, lines~\ref{alg2:phase2_start}-\ref{alg2:phase2_end}), which aims at re-allocating of the entire flow $f(t,i,j)$. Candidate routing paths $p \in P(i,j)$ are evaluated from the shortest to the longer one. For a~particular path, a~special procedure tries to find first-fit free channel accumulating entire bit-rate. If the full re-allocation is possible, all existing light-paths are deleted and their resources are released. Then, a~new light-path $l^\prime = (c,p \in P(i,j))$ is established. Note that $b_{cur}(l^\prime) = f(t,i,j)$. When re-allocation of the entire flow is infeasible, the method applies third strategy (Alg.~\ref{alg:increased_traffic}, lines~\ref{alg2:phase3_start}-\ref{alg2:phase3_end}), which implements partial flow re-allocation. In more detail, the method tries to re-allocate new flow with one of the existing light-paths. The light-paths are considered from the newest to the oldest and for each $l \in L(i,j)$ the method goes through paths $p \in P(i,j)$ (starting from the shortest one) in order to create a~light-path $l^\prime = (c, p \in P(i,j))$ accommodating $f(t,i,j) - b_{cur}(i,j) + b_{cur}(l)$. If such a~light-path is found, light-path \textit{l} is deleted and $l^\prime$ is established. The method then terminates. If the third strategy fails, the process moves to the next one -- addition to existing light-paths (Alg.~\ref{alg:increased_traffic}, lines~\ref{alg2:phase4_start}-\ref{alg2:phase4_end}). It tries to divide the offered bit-rate and add it to existing light-paths, which are considered from the oldest to the newest one. If it succeeds, the method terminates. Otherwise, it tries to establish a~new light-path accommodating still unallocated flow (Alg.~\ref{alg:increased_traffic}, lines~\ref{alg2:phase5_start}-\ref{alg2:phase5_end}). For that purpose, candidate paths $p \in P(i,j)$ are analyzed starting from the shortest to the longest one. If the allocation is not possible, the rest of flow is rejected.

\begin{algorithm}
\caption{$\textsc{tdrsa\_higher\_flow}(...)$ function}\label{alg:increased_traffic}

\begin{small}

\DontPrintSemicolon
\KwInput{$(i,j)$, $P(i,j)$, $L(i,j)$, $f(t,i,j)$, $b_{offered}(t,i,j)$, $b_{rej}$}

\tcc{There are not already allocated light-paths for $(i,j)$}
\If{$L(i,j) = \emptyset$}
{
    \For{$p \in P(i,j)$}
    {
         $c \gets find\_ff\_channel(p, f(t,i,j))$
         
         \If{$c \neq \emptyset$}
         {
            allocate allocate bit-rate $f(t,i,j)$ using light-path $l = (c, p)$
            
            $L(i,j) = L(i,j) \cup l$
            
            \textbf{return}
         }
    }
    
    $b_{rej}$ = $b_{rej}$ + $f(t,i,j)$
    
    \textbf{return}
}

\tcc{There are already allocated light-paths for $(i,j)$}

\Else
{
    \tcp*{Add to an existing light-path}
    \For{$l \in L(i,j)$} 
    {\label{alg2:phase1_start}
        \If{$b_{offered}(t,i,j) \leq b_{max}(l) - b_{cur}(l)$}
        {
             $b_{cur}(l) = b_{cur}(l) + b_{offered}(t,i,j)$
             
             \textbf{return}
        }
    }\label{alg2:phase1_end}
    
    \tcp*{Re-allocate the entire flow}
    \For{$p \in P(i,j)$}
    {\label{alg2:phase2_start}
        $c \gets find\_ff\_channel(p,f(t,i,j)$
        
        \If{$c \neq \emptyset$}
        {
            allocate bit-rate $f(t,i,j)$ using light-path $l^\prime=(c,p)$
            
            $L(i,j) = L(i,j) \cup l^\prime$
            
            \textbf{return}
        }
    }\label{alg2:phase2_end}
    
    \tcp*{Re-allocate a~part of flow}
    \For{$l \in L(i,j)$}
    {\label{alg2:phase3_start}
        \For{$p \in P(i,j)$}
        {
             $c \gets find\_ff\_channel(p,b_{cur}(l) + b_{offered}(t,i,j))$
             
             \If{$c \neq \emptyset$}
             {
                $L(i,j) = L(i,j) \setminus l$
                
                allocate bit-rate $b_{cur}(l)+b_{offered}(t,i,j)$ using light-path $l^\prime=(c,p)$
                
                $L(i,j) = L(i,j) \cup l^\prime$
                
                \textbf{return}
             }
        }
    }\label{alg2:phase3_end}
    
    \tcp*{Add to existing light-paths}
    \For{$l \in L(i,j)$}
    {\label{alg2:phase4_start}
        \If{$b_{cur}(l) < b_{max}(l)$}
        {
            \If{$b_{offered}(t,i,j) \leq b_{max}(l) - b_{cur}(l)$}
            {
                $b_{cur}(l) = b_{cur}(l) + b_{offered}(t,i,j)$
                
                \textbf{return}
            }
            
            \Else
            {
                $b_{offered}(t,i,j) = b_{offered}(t,i,j) - (b_{max}(l) - b_{cur}(l))$
                
                $b_{cur}(l) = b_{max}(l)$
            }
        }
    }\label{alg2:phase4_end}
    
    \tcp*{Allocate rest of the flow}
    \For{$p \in P(i,j)$}
    {\label{alg2:phase5_start}
        $c \gets find\_ff\_channel(b)$
        
        \If{$c \neq \emptyset$}
        {
            allocate bit-rate $b_{offered}(t,i,j)$ using light-path $l^\prime=(c,p)$
            
            $L(i,j) = L(i,j) \cup l^\prime$
             
            \textbf{return}
        }
    }\label{alg2:phase5_end}

    $b_{rej} = b_{rej} + b_{offered}(t,i,j)$
    
    \textbf{return}
}

\end{small}

\end{algorithm}

\subsection{$\textsc{tdrsa\_lower\_flow}(...)$ function}

If the currently observed flow $f(t,i,j)$ is lower that the already allocated flow $b_{cur}(i,j)$, then the $\textsc{tdrsa\_lower\_flow}(...)$ function is called. Its main idea, which is presented in Alg.~\ref{alg:decreased_traffic}, assures decrease of the currently allocated flow starting from the youngest light-paths. Note that the bit-rate to be released is equal to $b_{torelease} = b_{cur}(i,j) - f(t,i,j)$. Light-paths, for which $b_{cur}(l) = 0$ after that operation, are deleted and their resources are free. 

\begin{algorithm}
\caption{$\textsc{tdrsa\_lower\_flow}(...)$ function}\label{alg:decreased_traffic}

\begin{small}

\DontPrintSemicolon

\KwInput{$L(i,j), b_{torelease}$}

\While{$b_{torelease}$}
{
    \For{$l \in L(i,j)$}
    {
        \If{$b_{cur}(l) \leq b_{torealease}$}
        {
            $b_{torelease} = b_{torelease} - b_{cur}(l)$
            
            $L(i,j) = L(i,j,) \setminus l$
        }
        
        \Else
        {
            $b_{cur}(l) = b_{cur}(l) - b_{torelease}$
            
            \textbf{return}
        }
    }
}

\end{small}
\end{algorithm}

\section{Service relocation policies}\label{sec:relocation-policies}

The idea of the service relocation process is to change the assigned ~\textsc{dc} for a~selected client node (i.e., non-\textsc{dc} city node). The procedure is performed in order to improve network performance in terms of the ratio of accepted traffic. Therefore, it consist of two phases: (\textit{i}) \textsc{dc}s selection and (\textit{ii}) service (client) selection. Based on the analysis of the current network performance, the first phases selects a~pair of \textsc{dc}s $(r_1,r_2): r_1,r_2 \in R; r_1 \neq r_2$ which will be subject to a~change. In more detail, one client currently served by $r_1$ will be moved to $r_2$. The second phase selects then a~service (which is related here to a~non-\textsc{dc} city node) for which selected \textsc{dc} will be changed from $r_1$ into $r_2$. 

We propose 3 different methods for the \textsc{dc}s selection phase and 7 for the service selection step. Since each method of the first phase can be combined with any of the approaches proposed for the second step, 21 different relocation policies are analysed. We denote them as $M\_Phase1/M\_Phase2$ wherein $M\_Phase1$ refers to the method used for \textsc{dc}s selection and $M\_Phase2$ is a~method for service selection. 

Four parameters control the relocation process -- $t_{start}$, $\alpha$, $\beta_{r}$ and $\beta_{t}$. $t_{start}$ indicates minimum history size (measured in the number of processed iterations) required for effective service relocation. $\alpha$ indicates interval (measured in the number of iterations) between two subsequent runs of relocation. $\beta_{r}$ and $\beta_{t}$ refer to the precision used to indicate \textsc{dc}s for relocation. Parameters $t_{start}$ and $\alpha$ are utilized in all relocation policies. $p_{r}$ applies to \textsc{rb} and \textsc{h} while $p_{t}$ refers to \textsc{tb} and \textsc{h}. 

\subsection{\textsc{dc}s selection methods}

Three methods are proposed for \textsc{dc}s selection phase. Depending on the mechanism used for the decision making, they are called: rejection-based (\textsc{rb}), traffic-based (\textsc{tb}) and hybrid (\textsc{h}). The methods use historical data (i.e., observed bit-rate rejection and offered traffic volume) gathered during a~given history window. If the relocation process is run for the first time, the window includes all already processed iterations (i.e., $t_{start}$). Otherwise, the history takes into account the period since the last relocation process (i.e., $\alpha$ iterations).

The rejection-based strategy analysis rejected bit-rate in a~given history window. In particular, it sums bit-rate rejected within the window for each \textsc{dc} node (i.e., bit-rate sent from or to that \textsc{dc}). It also determines total rejected bit-rate for nodes hosting \textsc{dc}s (it is a~sum of rejected bit-rated obtained for each \textsc{dc} node). Then, it selects the pair of \textsc{dc}s with the highest (among all pairs of \textsc{dc}s) absolute value of the rejection difference. If that value it higher that $\beta_{r}$ of total rejected bit-rate, that pair of \textsc{dc}s is selected for the relocation. Otherwise, no relocation is performed at the moment.

The traffic-based method utilizes flow history to build auto-regressive integrated moving average (\textsc{arima}) model for traffic forecasting. It uses that model to predict total traffic volume for each \textsc{dc} node for the period until the next attempt of the relocation (i.e., $\alpha$ iterations). Then, it selects pair of \textsc{dc}s with the highest (among all pairs of \textsc{dc}s) absolute value of the prediction difference. If the value it higher that $\beta_{t}$ of total predicted bit-rate (for all \textsc{dc} nodes), the pair is selected for the change. Otherwise, no relocation is performed at the moment. 

Eventually, the hybrid strategy benefits from both -- rejection history and traffic prediction. It determines two groups of candidate pairs of \textsc{dc}s for relocation. The first group is calculated based on the rejection history -- it contains pairs for which the absolute value of the rejection difference is higher than $\beta_{r}$ of total rejected bit-rate. The second group consists of pairs for which the absolute value of the traffic prediction is higher than $\beta_{t}$ of total predicted bit-rate. Then, the strategy considers only pairs which are included in both groups and finally selects the pair for which sum of $\beta_{r}$ multiplied by the total rejection and $\beta_{p}$ multiplied by the total prediction is the highest. If two groups of candidate pairs do not have any common elements, there is no relocation at the moment. 

\subsection{Service selection methods}

Having selected a~pair of \textsc{dc}s for relocation $(r_1,r_2)$, the next step is to determine a~service (client node) that will be relocated from $r_1$ to $r_2$. The candidate nodes for relocation are nodes currently served by $r_1$. To select one of them, we propose 7 different policies: Rand, MinD, MaxD, MinR, MaxR, MinT and MaxT. 

Rand policy is the simplest one and it relocates a~randomly selected candidate. The next two policies (i.e., MinD, MaxD) make decisions based on the lengths (in kilometres) of shortest paths connecting candidate nodes with $r_1$ and $r_2$. MinD strategy chooses the candidate that is the closest to $r_2$ while MaxD policy selects the most distant client from $r_1$. Then, policies MinR and MaxR relocate the node with, accordingly, minimum or maximum rejected bit-rate observed within given history window. Eventually, MinT and MaxT analyze traffic forecast for each of the candidates and select the node with, respectively, lowest or highest prediction. 

\section{Numerical experiments}

This section discusses numerical experiments focused on three aspects: (\textit{i})~tuning of the proposed relocation policies, (\textit{ii})~their comparison and selection of best one, (\textit{iii})~case study -- benefits provided by the service relocation process in dynamic and cloud-ready \textsc{eon}s. 

\subsection{Simulation setup}

\begin{figure}[ht]
	\centering
	\includegraphics[width=0.9\linewidth]{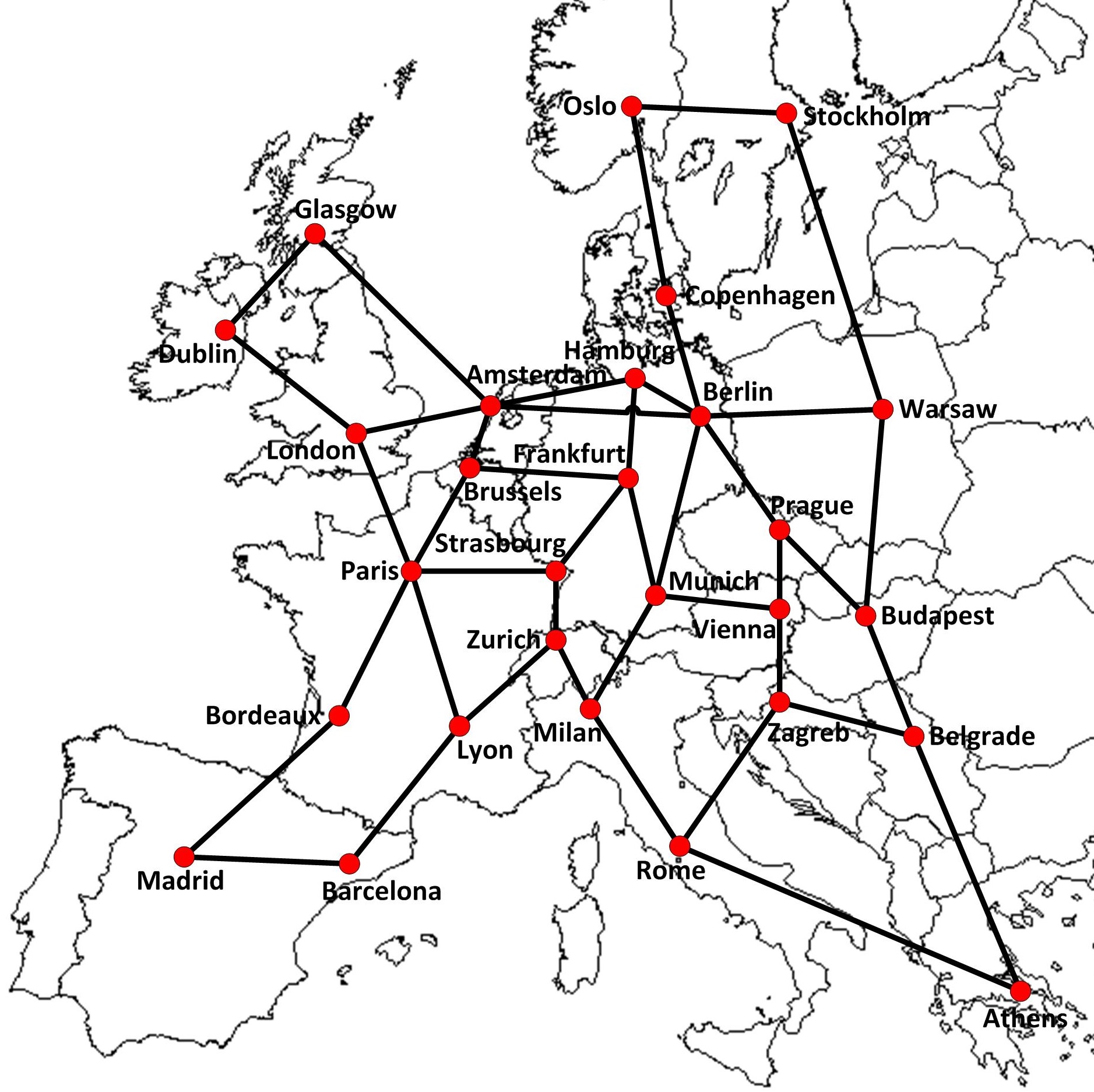}
	\caption{Euro28 network topology}
	\label{fig:euro28}	
\end{figure}

In all experiments, we consider Euro28 topology (28 nodes, 62 links of 625 km average length), which models European national network. The topology is depictured in Fig.~\ref{fig:euro28}. We assume that number of available \textsc{dc}s is $|R| \in \{ 3, 5, 7, 9, 11 \}$ and they are placed based on realistic data provided by https://www.datacentermap.com/. Table~\ref{tab:dcs} summarizes nodes hosting \textsc{dc}s for various number of \textsc{dc}s.

\begin{table}[htb]
    \centering
    \caption{Location of data centers in Euro28 network}
    \label{tab:dcs}
    %\begin{footnotesize}
\begin{tabular}{p{0.15\linewidth} | p{0.75\linewidth}}
        
        & \textbf{cities hosting data centers } \\ \hline\hline
        $|R|=3$ & London, Paris, Amsterdam \\
        $|R|=5$ & London, Paris, Amsterdam, Zurich, Frankfurt \\
        $|R|=7$ & London, Paris, Amsterdam, Zurich, Frankfurt, Madrid, Warsaw \\
        $|R|=9$ & London, Paris, Amsterdam, Zurich, Frankfurt, Milan, Vienna, Madrid, Warsaw \\
        $|R|=11$ & London, Paris, Brussels, Amsterdam, Zurich, Frankfurt, Milan, Vienna, Madrid, Warsaw, Copenhagen \\
    
    \end{tabular}
    %\end{footnotesize}
\end{table}

For the purpose of traffic model, we assume average traffic volume to be $B_{avg} \in \{ 50, 51, 52, ..., 59, 60 \}$ Tbps and the network observation time perspective to be $T = 3000$ iterations. The values of \textsc{gdg} and \textsc{pop} parameters were gathered from the official websites of the cities related to Euro28 nodes at the end of 2020. The distances between cities reflect real geographical distances given in kilometres. 

We use similar network physical model as in \cite{Khodashenas_lighttech16}. We work under the assumption that a~transponder occupies 3 frequency slices (37.5~GHz) and can use one of four modulation formats: \textsc{bpsk}, \textsc{qpsk}, 16-\textsc{qam}, 32-\textsc{qam} (characterized in Table~\ref{tab:modulations}). When a~path length exceeds a~modulation transmission distance, we use signal regenerators. To select a~modulation for a~particular bit-rate and a~candidate routing path, (\textsc{dat}) rule is used \cite{Walkowiak_springer16}. It applies the most spectrally efficient format which minimizes the number of required regenerators at the same time.

The main comparison criterion in experiments is bandwidth blocking probability (\textsc{bbp}). In some cases we also compare processing time. Since the traffic model is not fully deterministic, we repeat calculations for each scenario 5 times and presents averaged results. Due to the limited space of that paper, we present results for selected configurations, which present the general observed trends. 

[calculations]

\subsection{Tuning of the algorithms}

First step of the experiments is focused on the tuning of the proposed \textsc{tdrsa} algorithm and relocation policies. There are overall 21 different policies taking into account 3 methods of \textsc{dc}s selection phase (i.e., \textsc{rb}, \textsc{TB}, \textsc{H}) and 7 for the client selection phase (i.e.., \textsc{Rand}, \textsc{MinR}, \textsc{MaxR}, \textsc{MinT}, \textsc{MaxT}, \textsc{MinD}, \textsc{MaxD}). 

\textsc{tdrsa} algorithm has one input parameter $\lambda$, which defines the depth of the light-path searching process for each bit-rate request. In mode detail, it determines number of candidate routing paths that are taken into account while creating candidate light-paths. Based on our previous experiments, we consider $\lambda \in \{ 5, 10, 15, 20, 25, 30 \}$. The input parameters of the relocation policies are: $t_{start}$, $\alpha$, $\beta_{r}$ and $\beta_{t}$. Based on the initial experiments, we assume that $t_{start}=300$ and we consider following values of the rest of parameters: $\alpha \in \{ 50, 100, 150, ... 450, 500 \}$, $\beta_{r}, \beta_{t} \in \{ 0.05, 0.1, ..., 0.45, 0.5 \}$. In the tuning, we assume $B_{avg}=55$ Tbps and $|R|=7$. 

\begin{figure*}[ht]
	\centering
	\includegraphics[width=0.75\linewidth]{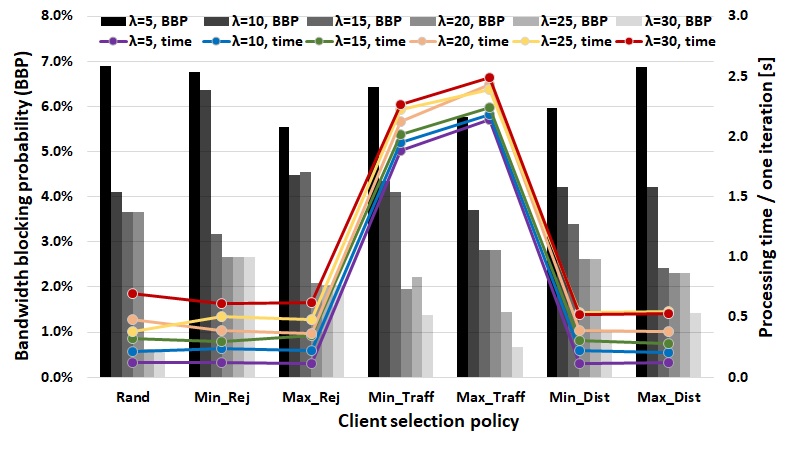}
	\caption{Tuning of \textsc{tdrsa} algorithm and relocation policies -- average \textsc{bbp} as~function of $\lambda$ for $B_{avg}=55$ Tbps and $R=7$}
	\label{fig:alg_x1}	
\end{figure*}

In our investigation we first tune relocation policies using $\lambda=5$ and next select best $\lambda$ value for the general allocation algorithm. To this end, for each policy we simulate network operation (and obtain average \textsc{bbp}) considering all possible combinations of parameters $\alpha$, $\beta_{r}, \beta_{t}$ and finally select the combination providing the smallest blocking probability. Table~\ref{tab:tuning} reports best configurations revealed for each policy. Then, we use best found configurations (i.e., Table~\ref{tab:tuning}) to simulate network operation for various values of $\lambda$ parameter and in turn to tune \textsc{tdrsa} algorithm. The results show that this parameter influences the most efficiency of the proposed approaches. Fig.~\ref{fig:alg_x1} reports dependencies obtained for \textsc{rb} policy and shows that $\lambda$ impacts significantly methods' performance in terms of two relevant criteria -- \textsc{bbp} and time of calculations. The higher $\lambda$ value, the lower blocking probability. However, at the price of longer calculations. For instance, the application of $\lambda=30$ instead of $\lambda=5$ for \textsc{rb/Rand} allowed to served 6.27\% more traffic, which is on average equal to 3.45 Tbps. Alongside the processing time has increased nearly 5.6-fold. Note that calculations of policies implementing traffic forecasting (in this case -- \textsc{MinT} and \textsc{MaxT} are several times longer than the calculations of other policies, which prove high complexity of the prediction task. However, the processing times of all policies are acceptably short and therefore we recommend to use $\lambda=30$ for further experiments. 

\begin{table*}[th]
    \centering
    \caption{Tuning of the service relocation policies -- recommended values of $\alpha | \beta_{r} | \beta_{t}$ parameters}
    \label{tab:tuning}
	\begin{tabular}{c||ccccccc} \
& \textsc{Rand} & \textsc{MinR} & \textsc{MaxR} & \textsc{MinT} & \textsc{MaxT} & \textsc{MinD} & \textsc{MaxD}  \\
\hline \hline
\textsc{rb} & 250 $|$ 0.4 $|$ - & 50 $|$ 0.25 $|$ - & 450 $|$ 0.15 $|$ - & 450 $|$ 0.45 $|$ - & 50 $|$ 0.35 $|$ - & 350 $|$ 0.2 $|$ - & 450 $|$ 0.35 $|$ - \\
\textsc{tb} & 150 $|$ - $|$ 0.4 & 250 $|$ - $|$ 0.15 & 200 $|$ - $|$ 0.4 & 200 $|$ - $|$ 0.2 & 450 $|$ - $|$ 0.45 & 150 $|$ - $|$ 0.4 & 450 $|$ - $|$ 0.55 \\
\textsc{h} & 200 $|$ 0.15 $|$ 0.2 & 450 $|$ 0.15 $|$ 0.15 & 50 $|$ 0.05 $|$ 0.25 & 100 $|$ 0.15 $|$ 0.05 & 200 $|$ 0.1 $|$ 0.1 & 250 $|$ 0.15 $|$ 0.05 & 450 $|$ 0.2 $|$ 0.2 \\
%\textsc{rb} & 250 \mid 0.4 \mid - & 50 \mid 0.25 \mid - & 450 \mid 0.15 \mid - & 450 \mid 0.45 \mid - & 50 \mid 0.35 \mid - & 350 \mid 0.2 \mid - & 450 \mid 0.35 \mid - \\
%\textsc{tb} & 150 \mid - \mid 0.4 & 250 \mid - \mid 0.15 & 200 \mid - \mid 0.4 & 200 \mid - \mid 0.2 & 450 \mid - \mid 0.45 & 150 \mid - \mid 0.4 & 450 \mid - \mid 0.55 \\
%\textsc{h} & 200 \mid 0.15 \mid 0.2 & 450 \mid 0.15 \mid 0.15 & 50 \mid 0.05 \mid 0.25 & 100 \mid 0.15 \mid 0.05 & 200 \mid 0.1 \mid 0.1 & 250 \mid 0.15 \mid 0.05 & 450 \mid 0.2 \mid 0.2 \\
\end{tabular}
\end{table*}

\subsection{Comparison of service relocation policies}

\begin{figure*}[ht]
	\centering
	\includegraphics[width=0.75\linewidth]{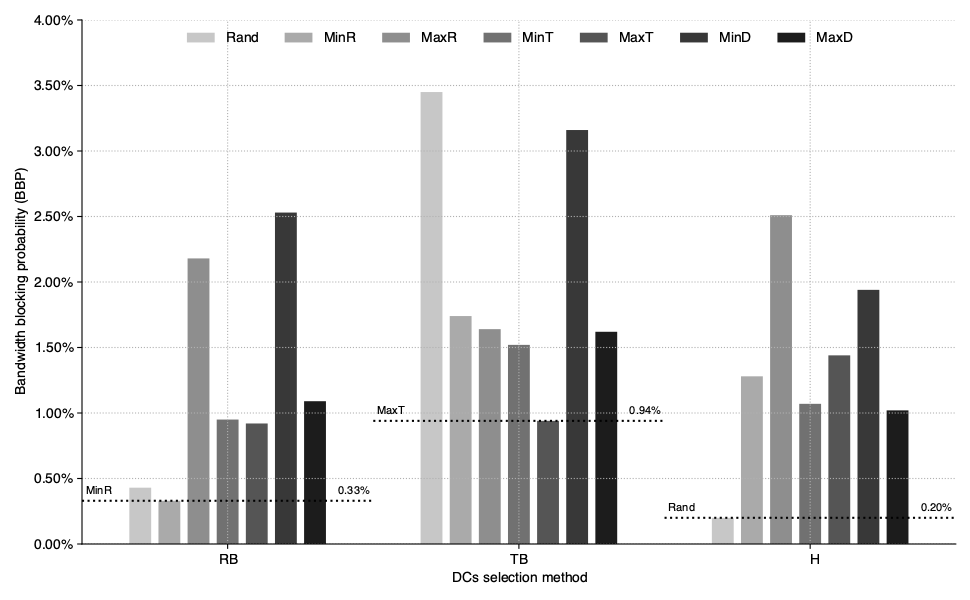}
	\caption{Comparison of service relocation policies -- average \textsc{bbp} for $B_{avg}=55$ Tbps, $R=7$, $\lambda=30$}
	\label{fig:alg_cmpfig}	
\end{figure*}

The next step of the experiments is comparison of proposed service relocation policies. To this end, we simulate network operation using the tuned allocation algorithm and relocation policies (i.e., polices with parameters set according to Table~\ref{tab:tuning}) and present averaged results in~Fig.~\ref{fig:alg_cmpfig}. The findings show that the client selection method has a~significant impact on the relocation process efficiency. For instance, considering method \textsc{h} for the \textsc{dc}s selection phase, we can serve 2.31\% more traffic when applying \textsc{Rand} instead of \textsc{MaxR} approach in the client selection phase. Similarly for \textsc{rb} and \textsc{tb} policies, the differences between results of best and worst client selection method are equal to, respectively, 2.19\% and 2.51\%. The experiments show that the \textsc{tb} performs the best combined with \textsc{MaxT}, the \textsc{rb} together with \textsc{MinR} while the \textsc{h} provides best results when applied with \textsc{Rand} policy. Hence, when the \textsc{dc}s selection is based on a~one data type (i.e., rejection history or traffic prediction), it is beneficial to use the same data type in the client selection phase. When the first phase of the relocation process makes use of two data types, then it is recommended to randomly select client node to be relocated.

It is worth-mentioning that \textsc{h/Rand} policy performed the best among all tested policies in the experiments and allowed to allocate, respectively, 0.13\% and 0.74\% more traffic than \textsc{rb/MinR} and \textsc{tb/MaxT} policies.    

\section{Conclusions}\label{sec:conclusions}

In this paper, we have focuses on the problem of efficient service relocation in elastic optical networks in order to increase network performance (measured by the volume of accepted traffic). To this end, we first propose novel traffic model for cloud ready transport networks. The model takes into account four flow types (i.e., city-to-city, city-to-data center, data center-to-data center and data center-to-data center) while the flow characteristics are based on real economical and geographical parameters of the cities related to network nodes. Due to the specific definition of the traffic model, we propose dedicated flow allocation algorithm (i.e., \textsc{tdrsa}). To achieve high efficiency and fast calculations, \textsc{tdrsa} aims at minimizing number of established light-paths and (re-)allocation actions. Its process can be additionally improved by the application of the service relocation process. For that purpose, we introduce 21 different relocation policies, which use three types of data for decision making -- network topological characteristics, rejection history and traffic prediction. Eventually, we perform extensive simulations in order to: (\textit{i}) tune proposed optimization approaches and (\textit{ii}) evaluate and compare their efficiency and select the best one. The results of the numerical experiments prove high efficiency of the proposed relocation process. The propoerly designed relocation policy allowed to serve up to 3\% more traffic compared to the network operation without that policy. They also reveal that the most efficient relocation policy bases its decisions on two types of data simultaneously -- the rejection history and traffic prediction.

In the future work, we plan to extend our investigation to survivable networks, in which rejection history and traffic predictions will be used to provide or improve network resilience to failures and attacks. 

\section*{Acknowledgment}
This work was supported by the Polish National Science Centre (NCN) under Grant DEC-2018/31/D/ST6/03041.

\ifCLASSOPTIONcaptionsoff
  \newpage
\fi

%\FloatBarrier

\bibliographystyle{IEEEtran}
\bibliography{goscien_reloc.bib}


\end{document}